\documentclass[english, useAMS, usenatbib, usegraphicx]{mn2e}
\usepackage[T1]{fontenc}
\usepackage[latin9]{inputenc}
\usepackage{amsmath}
\usepackage{esint}
\usepackage{babel}
\begin{document}

\title{Future deceleration due to cosmic backreaction in presence of the 
event horizon}

\author[N. Bose, A. S. Majumdar]{Nilok Bose\thanks{Email: nilok@bose.res.in}, A. S. Majumdar\thanks{Email: archan@bose.res.in} \\
S. N. Bose National Centre for Basic Sciences, Block JD, Sector III, Salt Lake, Calcutta 700098, India}

\date{Released 2011}

\maketitle

\begin{abstract}
The present acceleration of the Universe leads to the formation of
a cosmological future event horizon. We explore the effects of the
event horizon on cosmological backreaction due to inhomogeneities
in the universe. Beginning from the onset of the present accelerating
era, we show that backreaction in presence of the event horizon causes
acceleration to slow down in the subsequent evolution. Transition
to another decelerating era could ensue eventually at a future epoch, ensuring 
avoidance of a big rip.
\end{abstract}

\section{Introduction}

There exists overwhelming observational evidence for the present acceleration
of the Universe \citep{perlmutter, kowalski, hicken, schwarz}. The accelerating universe leads to
a future event horizon from beyond which it is not possible for any
signal to reach us. On the other hand, observations also tell us that
our Universe is inhomogeneous up to the scales of super clusters
of galaxies. The idea that backreaction originating from density
inhomogeneities could lead to modifications in evolution of the universe as
described by the background Friedmann-Robertson-Walker (FRW) metric at large
scales has gained popularity in recent years \citep{buchert-1,buchert-2,buchert-3,zalaletdinov-1,buchert-4,rasanen-1,rasanen-2,rasanen-3,kolb-1,kolb-2,paranjape-1,singh,mattson,wiltshire-1,wiltshire-2,gasperini}.
Here we show
that backreaction in the presence of the cosmological event horizon could
have a remarkable consequence of ushering in another decelerated era beyond
the present accelerating epoch.

In spite of numerous creative ideas proposed for
the present acceleration \citep{sahni,copeland},
there is still a lack of convincing explanation of this
phenomenon.  The simplest possible explanation
 provided by a cosmological constant
is endowed with conceptual problems \citep{weinberg}.
Alternative mechanisms based on either modifications of the gravitational
theory, or invoking extra fields
suffer from the coincidence problem, as to why the era
of acceleration begins around the same era when the Universe
becomes structured. The ultimate fate of our Universe
remains clouded in considerable mystery. Backreaction from inhomogeneities
provides an interesting platform for investigating this issue without
invoking additional physics, since
the effects of backreaction gain stength as the inhomogeneities develop
into structures around the present era.

Approaches have been developed to calculate the effect of
inhomegeneous matter distibution  on the
evolution of the Universe
\citep{buchert-1,zalaletdinov-1,kolb-1}. Arguments
in favour of the viability of backreaction seem rather compelling
\citep{kolb-2}, though there exists debate on the impact of
inhomogeneities on observables of an overall homogeneous FRW model
\citep{paranjape-1,wald,singh}, and on the magnitude of
backreaction modulated by the effect of shear between overdense and
underdense regions \citep{mattson}. Using the framework fomulated
by Buchert \citep{buchert-1}
it has been shown \citep{rasanen-1,rasanen-2,rasanen-3} that backreaction could
lead to an accelerated expansion during the present epoch.
A notable application of the formalism has been developed
by Wiltshire showing an apparent volume acceleration of the universe
based on the different lapse of time between the underdense and overdense
regions \citep{wiltshire-1,wiltshire-2}. Further, gauge invariant
averages in the Buchert framework have also been constructed recently
\citep{gasperini}.

While upcoming observations may ultimately decide whether
backreaction from density inhomogeneities drives
the present acceleration, the above studies \citep{buchert-1,buchert-2,buchert-3,buchert-4,rasanen-1,rasanen-2,rasanen-3,kolb-1,kolb-2,wiltshire-1,wiltshire-2,gasperini} have highlighted that
backreaction could be an important ingredient of the evolution of our Universe.
Here we explore this issue
with a fresh perspective, \textit{viz.}, the impact of the event horizon
on cosmological backreaction.
The currently accelerating epoch dictates the existence of an event
horizon since the transition from the previously matter dominated
decelerating expansion. Since backreaction is evaluated from
the global distribution of matter inhomogeneities, the event horizon
demarcates the spatial regions which are causally connected to us and
hence impact the evolution of our part of the Universe.  In the present
work we investigate the consequences of backreaction in presence of 
the horizon. Such
an approach has remained unexplored in previous studies of backreaction.
It may be noted that the formalism of back reaction 
\citep{buchert-1,buchert-2,buchert-3} has been criticized on the grounds
that the average is taken on a space like
hypersurface, while observations are made along and inside the past light 
cone \cite{wald}. Our present analysis, by considering an effect due to the 
event horizon, introduces an element of light cone physics from a somewhat 
different perspective.
We show that backreaction with the event horizon could lead to the 
possibility of transition to a decelerated future era.

\section{The Backreaction Framework}

In the framework developed by Buchert \citep{buchert-1,buchert-2,buchert-3} for
 the Universe filled with an irrotational fluid of dust the
spacetime is foliated into flow\textendash{}orthogonal hypersurfaces
featuring the line\textendash{}element
$ds^{2}=-dt^{2}+g_{ij}dX^{i}dX^{j}$,
 where the proper time $t$ labels the hypersurfaces and $X^{i}$
are Gaussian normal coordinates (locating free\textendash{}falling
fluid elements or generalized fundamental observers) in the hypersurfaces,
and $g^{ij}$ is the full inhomogeneous three metric of the hypersurfaces
of constant proper time. For a compact spatial domain
$\mathcal{D}$ whose volume is given by $|\mathcal{D}|_{g}=\int_{\mathcal{D}}d\mu_{g}$
where $d\mu_{g}=\sqrt{^{(3)}g(t,X^{1},X^{2},X^{3})}dX^{1}dX^{2}dX^{3}$,
the scale factor $a_{\mathcal{D}}(t)=\left(\frac{|\mathcal{D}|_{g}}{|\mathcal{D}_{i}|_{g}}\right)^{1/3}$
encodes the average stretch of all directions of the domain. The
 Einstein equations then lead to
\citep{buchert-1,buchert-2,buchert-3}
\begin{eqnarray}
3\frac{\ddot{a}_{\mathcal{D}}}{a_{\mathcal{D}}} & = & -4\pi G\left\langle \rho\right\rangle _{\mathcal{D}}+\mathcal{Q}_{\mathcal{D}}+\Lambda\nonumber\\
3H_{\mathcal{D}}^{2} & = & 8\pi G\left\langle \rho\right\rangle _{\mathcal{D}}-\frac{1}{2}\mathcal{\left\langle R\right\rangle }_{\mathcal{D}}-\frac{1}{2}\mathcal{Q}_{\mathcal{D}}+\Lambda\label{eq:5}\\
0 & = & \partial_{t}\left\langle \rho\right\rangle _{\mathcal{D}}+3H_{\mathcal{D}}\left\langle \rho\right\rangle _{\mathcal{D}}\nonumber
\end{eqnarray}
where the average of the scalar quantities on the domain $\mathcal{D}$
is defined as
$\left\langle f\right\rangle {}_{\mathcal{D}}(t)=\frac{\int_{\mathcal{D}}f(t,X^{1},X^{2},X^{3})d\mu_{g}}{\int_{\mathcal{D}}d\mu_{g}}=|\mathcal{D}|_{g}^{-1}\int_{\mathcal{D}}fd\mu_{g}$,
 and where $\rho$, $\mathcal{R}$ and $H_{\mathcal{D}}$ denote the
local matter density, the Ricci-scalar of the three-metric $g_{ij}$,
and the domain dependent Hubble rate $H_{\mathcal{D}}=\dot{a}_{\mathcal{D}}/a_{\mathcal{D}}$
respectively. The kinematical backreaction $\mathcal{Q_{D}}$ is defined
as
$\mathcal{Q_{D}}=\frac{2}{3}\left(\left\langle \theta^{2}\right\rangle _{\mathcal{D}}-\left\langle \theta\right\rangle _{\mathcal{D}}^{2}\right)-2\sigma_{\mathcal{D}}^{2}$,
 where $\theta$ is the local expansion rate and $\sigma^{2}=1/2\sigma_{ij}\sigma^{ij}$
is the squared rate of shear. $\mathcal{Q_{D}}$ encodes the departure
from homogeneity.

The {}``global'' domain $\mathcal{D}$ is assumed to be separated
into subregions $\mathcal{F}_{\ell}$  which themselves consist of
elementary space entities $\mathcal{F}_{\ell}^{(\alpha)}$ that may
be associated with some averaging length scale, i.e.,
$\mathcal{D}=\cup_{\ell}\mathcal{F}_{\ell}$, where $\mathcal{F}_{\ell}=\cup_{\alpha}\mathcal{F}_{\ell}^{(\alpha)}$
and $\mathcal{F}_{\ell}^{(\alpha)}\cap\mathcal{F}_{m}^{(\beta)}=\emptyset$
for all $\alpha\neq\beta$ and $\ell\neq m$. Analogous to the scale factor for the global domain, a scale factor
$a_{\ell}$ for each of the subregions $\mathcal{F}_{\ell}$ can be
defined such that
$|\mathcal{D}|_{g}=\sum_{\ell}|\mathcal{F}_{\ell}|_{g}$, and hence
$a_{\mathcal{D}}^{3}=\sum_{\ell}\lambda_{\ell_{i}}a_{\ell}^{3}$,
where $\lambda_{\ell_{i}}=|\mathcal{F}_{\ell_{i}}|_{g}/|\mathcal{D}_{i}|_{g}$
is the initial volume fraction of the subregion $\mathcal{F}_{\ell}$.
The average of the scalar
valued function $f$ on the domain $\mathcal{D}$,
may then be split into the averages of $f$ on the subregions $\mathcal{F}_{\ell}$
in the form, $\left\langle f\right\rangle _{\mathcal{D}}=\underset{\ell}{\sum}|\mathcal{D}|_{g}^{-1}\underset{\alpha}{\sum}\int_{\mathcal{F}_{\ell}^{(\alpha)}}fd\mu_{g}=\underset{\ell}{\sum}\lambda_{\ell}\left\langle f\right\rangle _{\mathcal{F}_{\ell}}$,
where $\lambda_{\ell}=|\mathcal{F}_{\ell}|_{g}/|\mathcal{D}|_{g}$,
is the volume fraction of the subregion $\mathcal{F}_{\ell}$.
Due to the $\left\langle \theta\right\rangle _{\mathcal{D}}^{2}$
term, the expression for the backreaction
$\mathcal{Q_{D}}$ is given by
 \begin{equation}
\mathcal{Q_{D}}=\underset{\ell}{\mathcal{\sum}}\lambda_{\ell}\mathcal{Q}_{\ell}+3\underset{\ell\neq m}{\sum}\lambda_{\ell}\lambda_{m}\left(H_{\ell}-H_{m}\right)^{2}\label{eq:11}
\end{equation}
 where, $\mathcal{Q}_{\ell}$ and $H_{\ell}$ are defined in 
$\mathcal{F}_{\ell}$ in the same way as $\mathcal{Q_{D}}$ and $H_{\mathcal{D}}$ are defined in
$\mathcal{D}$. The shear part $\left\langle \sigma^{2}\right\rangle _{\mathcal{F}_{\ell}}$
is completely absorbed in $\mathcal{Q}_{\ell}$ whereas the variance
of the local expansion rates $\left\langle \theta^{2}\right\rangle _{\mathcal{D}}-\left\langle \theta\right\rangle _{\mathcal{D}}^{2}$
is partly contained in $\mathcal{Q}_{\ell}$ but also generates the
extra term $3\sum_{\ell\neq m}\lambda_{\ell}\lambda_{m}\left(H_{\ell}-H_{m}\right)^{2}$. 
This is because the part of the variance that is present in $\mathcal{Q}_{\ell}$, namely $\left\langle \theta^{2}\right\rangle _{\mathcal{F}_{\ell}}-\left\langle \theta\right\rangle _{\mathcal{F}_{\ell}}^{2}$
only takes into account points inside $\mathcal{F}_{\ell}$. To restore the variance that comes from combining points of $\mathcal{F}_{\ell}$ with others in $\mathcal{F}_{m}$, the extra term containing the averaged Hubble rate emerges.
Note here that the above formulation of the backreaction holds
in the case when there is no interaction between the overdense and the
underdense subregions.

Now from Eq.\eqref{eq:5} one gets
\begin{equation}
\frac{\ddot{a}_{\mathcal{D}}}{a_{\mathcal{D}}}=\underset{\ell}{\sum}\lambda_{\ell}\frac{\ddot{a}_{\ell}(t)}{a_{\ell}(t)}+\underset{\ell\neq m}{\sum}\lambda_{\ell}\lambda_{m}\left(H_{\ell}-H_{m}\right)^{2}\label{eq:13}
\end{equation}
Following the simplifying assumption of Ref.\citep{buchert-3},
(which captures the essential physics) we work with only two subregions.
Clubbing those parts of $\mathcal{D}$ which consist of initial overdensity
as $\mathcal{M}$ (called {}``wall''), and those with initial underdensity
as $\mathcal{E}$ (called {}``void''), such that $\mathcal{D}=\mathcal{M}\cup\mathcal{E}$,
one obtains $H_{\mathcal{D}}=\lambda_{\mathcal{M}}H_{\mathcal{M}}+\lambda_{\mathcal{E}}H_{\mathcal{E}}$,
with similar expressions for $\left\langle \rho\right\rangle _{\mathcal{D}}$
and $\left\langle \mathcal{R}\right\rangle _{\mathcal{D}}$, and
\begin{equation}
\frac{\ddot{a}_{\mathcal{D}}}{a_{\mathcal{D}}}=\lambda_{\mathcal{M}}\frac{\ddot{a}_{\mathcal{M}}}{a_{\mathcal{M}}}+\lambda_{\mathcal{E}}\frac{\ddot{a}_{\mathcal{E}}}{a_{\mathcal{E}}}+2\lambda_{\mathcal{M}}\lambda_{\mathcal{E}}(H_{\mathcal{M}}-H_{\mathcal{E}})^{2}\label{eq:15}
\end{equation}
Here $\lambda_{\mathcal{M}}+\lambda_{\mathcal{E}}=1$, with
 $\lambda_{\mathcal{M}}=|\mathcal{M}|/|\mathcal{D}|$ and
$\lambda_{\mathcal{E}}=|\mathcal{E}|/|\mathcal{D}|$.
Since the global domain $\mathcal{D}$ is large enough for a scale
of homogeneity to be associated with it, one can write
$|\mathcal{D}|_{g}=\int_{\mathcal{D}}\sqrt{-g}\, d^{3}X \approx f(r)a_{F}^{3}(t)$,
where $f(r)$ is a function of the FRW comoving radial coordinate
$r$. It then follows that $a_{\mathcal{D}} \approx \left(\frac{f(r)}{|\mathcal{D}_{i}|_{g}}\right)^{1/3}a_{F}$, and hence, the volume average
scale factor $a_{\mathcal{D}}$ and the FRW scale factor
 $a_{F}$ are related by $a_{\mathcal{D}} \approx  c_{F}a_{F}$, where $c_{F}$ is constant
in time. Thus, $H_{F} \approx H_{\mathcal{D}}$,
where $H_{F}$ is the FRW Hubble parameter associated with $\mathcal{D}$.
Though in general $H_D$ and $H_F$ could  differ on even large
scales \citep{buchert-3}, the above approximation is valid for small metric
perturbations.

\section{Effect of event horizon}

We now come to the central issue of the paper, as to what happens
to the evolution of the universe once the present stage of acceleration sets
in. Note henceforth, we do not need to necessarily assume that
the acceleration is due to backreaction \citep{buchert-3,rasanen-1}.
For the purpose of
our present analysis, it suffices to consider the observed accelerated
phase of the universe \citep{schwarz} that could occur due to any of a variety of
mechanisms \citep{sahni,copeland}. Given that we are undergoing a stage of acceleration
since transition from an era of structure formation, our aim here is to
explore the subsequent evolution of
the Universe due to the effects of backreaction in presence of the cosmic
event horizon. Though in general, spatial and light cone
distances and corresponding accelerations could be different, as shown
explicitly in the framework of LTB models \citep{bolejko}, an approximation for
the event horizon which forms at the
onset of acceleration could be defined by
\begin{equation}
r_{h} = a_{\mathcal{D}}\int_{t}^{\infty}\frac{dt'}{a_{\mathcal{D}}(t')}\label{horizon}
\end{equation}
in the same spirit as $a_{\mathcal{D}} =  c_{F}a_{F}$. 

\begin{figure}
 \includegraphics[height=5.5cm]{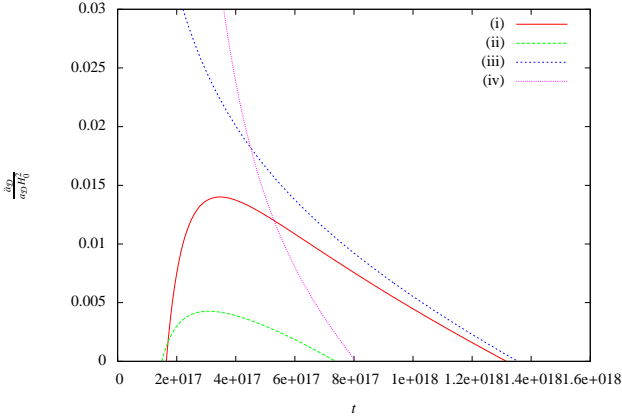} 
\caption{The dimensionless global acceleration parameter $\frac{\ddot{a}_{D}}{a_{D}H_{0}^{2}}$
is plotted versus time ($s$) assuming a constant horizon. The values for the various 
parameters used are (i) $\alpha=0.995,\beta=0.5$,
(ii) $\alpha=0.999,\beta=0.6$, (iii) $\alpha=1.0,\beta=0.5$, and
(iv) $\alpha=1.02,\beta=0.66$.}
\label{fig1}
\end{figure}

Following the Buchert framework \citep{buchert-1,buchert-3} as discussed
above, the
 global domain $\mathcal{D}$ is divided into a collection of overdense
regions $\mathcal{M}=\cup_{j}\mathcal{M}^{j}$,
with total volume $|\mathcal{M}|_{g}=\sum_{j}|\mathcal{M}^{j}|_{g}$, and
 underdense regions $\mathcal{E}=\cup_{j}\mathcal{E}^{j}$ with corresponding
volume $|\mathcal{E}|_{g}=\sum_{j}|\mathcal{E}^{j}|_{g}$.
Assuming that the scale factors of the regions $\mathcal{E}^{j}$
and  $\mathcal{M}^{j}$ are respectively given
by $a_{\mathcal{E}_{j}}=c_{\mathcal{E}_{j}}t^{\alpha}$ and $a_{\mathcal{M}_{j}}=c_{\mathcal{M}_{j}}t^{\beta}$,
where $\alpha$, $\beta$, $c_{\mathcal{E}_{j}}$ and $c_{\mathcal{M}_{j}}$
are constants, one has
 \begin{equation}
a_{\mathcal{E}}^{3}=c_{\mathcal{E}}^{3}t^{3\alpha};\>\>\>a_{\mathcal{M}}^{3}=c_{\mathcal{M}}^3t^{3\beta}\label{eq:19}
\end{equation}
 where $c_{\mathcal{E}}^{3}=\frac{\sum_{j}c_{\mathcal{E}_{j}}^{3}|\mathcal{E}_{i}^{j}|_{g}}{|\mathcal{E}_{i}|_{g}}$
is a constant, and similarly for $c_{\mathcal{M}}$.
The volume fraction of the subdomain $\mathcal{M}$ is given by $\lambda_{\mathcal{M}}=\frac{|\mathcal{M}|_{g}}{|\mathcal{D}|_{g}}$
which can be rewritten in terms of the corresponding scale factors
as $\lambda_{\mathcal{M}}=\frac{a_{\mathcal{M}}^{3}|\mathcal{M}_{i}|_{g}}{a_{\mathcal{D}}^{3}|\mathcal{D}_{i}|_{g}}$.
Since an event horizon forms, only those regions of $\mathcal{D}$
that are within the event horion are accessible to us. Hence, in this case
  an apparent volume fraction $\lambda_{\mathcal{M}_{h}}$
given by $\lambda_{\mathcal{M}_{h}}=\frac{a_{\mathcal{M}}^{3}|\mathcal{M}_{i}|_{g}}{\frac{4\pi}{3}r_h^{3}}$ is introduced.
From eq.(\ref{eq:19}) it follows that 
\begin{equation}
\lambda_{\mathcal{M}_{h}}=\frac{c_{\mathcal{M}_{h}}^{3}t^{3\beta}}{r_{h}^{3}}
\label{wallfrac}
\end{equation}
where $c_{\mathcal{M}_{h}}^{3}=3c_{\mathcal{M}}^{3}|\mathcal{M}_{i}|_{g}/4\pi$
is a constant. Normalizing the total
accessible volume in the presence of the event horizon, we can write
\begin{equation}
\lambda_{\mathcal{E}_{h}}=1-\lambda_{\mathcal{M}_{h}}
\label{voidfrac}
\end{equation}
 where $\lambda_{\mathcal{E}_{h}}$ is the apparent volume fraction
for the subdomain $\mathcal{E}$. It hence follows that the global acceleration
equation (\ref{eq:15}) is now given by
\begin{eqnarray}
\frac{\ddot{a}_{\mathcal{D}}}{a_{\mathcal{D}}} & = & \frac{c_{\mathcal{M}_{h}}^{3}t^{3\beta}}{r_{h}^{3}}\frac{\beta(\beta-1)}{t^{2}}+\left(1-\frac{c_{\mathcal{M}_{h}}^{3}t^{3\beta}}{r_{h}^{3}}\right)\frac{\alpha(\alpha-1)}{t^{2}}\nonumber \\
 &  & +2\frac{c_{\mathcal{M}_{h}}^{3}t^{3\beta}}{r_{h}^{3}}\left(1-\frac{c_{\mathcal{M}_{h}}^{3}t^{3\beta}}{r_{h}^{3}}\right)\left(\frac{\beta}{t}-\frac{\alpha}{t}\right)^{2}\label{eq:26}
\end{eqnarray}
In order to obtain the future evolution of the universe with backreaction
in presence of the event horizon, one has to solve the 
above equation for the scale factor with the event horizon $r_h$ given by 
Eq.(\ref{horizon}). In what follows we will eventually obtain numerical 
solutions of the above integro-differential equations. However, it is first 
instructive to obtain some physical insight of the evolution by taking 
recourse to a simple approximation. 

To this end, let us for the moment model the onset of the present acceleration 
of the Universe by an exponential expansion, keeping our analysis close to 
observations. Specifically, we set  $a_{\mathcal{D}}\propto e^{H_{\mathcal{D}}t}$
in Eq.(\ref{horizon}) only. (We will see later that this rather crude 
approximation does indeed give rise to results that are qualitatively similar 
to the ones obtained through numerical analysis).  
Using $H_{F}=H_{\mathcal{D}}$,
where $H_{F}$ is the FRW Hubble parameter associated with $\mathcal{D}$, it
follows that $r_{h}=H_{F}^{-1}$, a constant which we substitute in Eq.(\ref{eq:26}).
With this substitution, the  global 
acceleration $\ddot{a}_{\mathcal{D}}$ vanishes at times given by   
\begin{equation}
t^{3\beta}=\frac{r_{h}^{3}}{4\left(\beta-\alpha\right)c_{\mathcal{M}_{h}}^{3}}\left[\left(3\beta-\alpha-1\right)\\
\pm\sqrt{\left(3\beta-\alpha-1\right)^{2}+8\alpha\left(\alpha-1\right)}\right]\label{solnt}
\end{equation}
The scale factor of the {}``wall'' grows as $t^{\beta}$, where $1/2 \le \beta \le 2/3$.
Eq.\eqref{solnt}
corresponds to real time solutions for
$\alpha\geq\frac{1}{3}\left[\left(\beta+1\right)+2\sqrt{2\beta\left(1-\beta\right)}\right]\label{alphacond}$. 

Now, let us consider the following two cases separately: \textit{Case
I:} $\alpha<1$ and $\beta\le2/3$. There exist two real solutions
(\ref{solnt}) corresponding to two values of time when the global
acceleration vanishes. In Fig.1 we plot a dimensionless global acceleration
parameter $\frac{\ddot{a}_{D}}{a_{D}H_{0}^{2}}$ with time using eq.((\ref{eq:26})).
The curves (i) and (ii) correspond to this case showing that the Universe
first enters the epoch of acceleration due to backreaction, which
subsequently slows down and finally vanishes at the onset of another
decelerating era in the future. \textit{Case II:} $\alpha\geq1$ and
$\beta\le2/3$. From (\ref{solnt}) it follows that there is only
one real solution (minus sign for the square root). This case models
the Universe which accelerates due to some other mechanism (not backreaction),
but subsequently enters an epoch of deceleration due to backreaction
of inhomogeneities in the presence of the event horizon (see curves
(iii) and (iv) of Fig.1). 

The plots in Fig.1 have been done taking
the standard values of the parameters 
$r_{h}=H_{\mathcal{D}_{0}}^{-1}=4.36\times10^{17}s$
while choosing the appropriate range for the parameters $\alpha$
and $\beta$, as given in the figure caption. Based on the N-body
simulation values used in \citep{buchert-3} we also take $\lambda_{\mathcal{M}_{h0}}=0.09$.
Using the relation $z_{T}=\textrm{exp}\left[H_{\mathcal{D}_{0}}\left(t_{0}-t_{T}\right)\right]-1$,
where $t_{T}$ corresponds to the transition time in the past, the
red-shifts for the transition could be estimated. For example, for
the data used in curve (i), the transition from deceleration to acceleration
occurs at $z_{T}\simeq0.844$, and for curve (ii) we have $z_{T}\simeq0.914$
(which are close to the $\Lambda$CDM value for the standard transition
redshift \citep{melchiorri}).

We now study the acceleration equation (\ref{eq:26}) numerically
without assuming {\it a priori} any behaviour for the horizon. Keeping with the
spirit of our analysis, we assume that the Universe has entered the 
accelerated stage and thus a cosmic event horizon has formed. This ensures
that $r_h$ defined by (\ref{horizon}) will be finite valued, enabling us
to replace the integral equation (\ref{horizon}) by 
\begin{eqnarray}
\dot{r_{h}} & = & \frac{\dot{a}_{\mathcal{D}}}{a_{\mathcal{D}}}r_{h}-1\label{eq:28}
\end{eqnarray}
Thus, the evolution of the scale factor is now governed by the set of
coupled differential equations (\ref{eq:26}) and (\ref{eq:28}).
We numerically integrate these equations by using as an ``initial condition''
the  observational constraint  $q_{0}\approx-0.7$, where $q_{0}$
is the current value of the deceleration parameter, and using the solution
for the scale factor plot the global acceleration versus time in Fig.2 
(thus all the curves
in Fig.2 are set to intersect at the point $(t_0, q_0)$). The values
of the other parameters including $\alpha$ and $\beta$  are chosen to be
the same
as in the corresponding curves of the exponential case. 
As can be seen from Fig.2, the nature
of the plots is quite similar to the ones that were obtained in the
case assuming a constant event horizon, with the $\alpha > 1$ curves
signifying only one transition between acceleration and deceleration in 
the future. 
The differences in the various slopes and also
in the scale for the dimensionless global acceleration parameter in the two
cases arise as a result of the approximation of constant horizon used in the
former, as well as due to
the choice of the condition $q_{0}\approx-0.7$ used in the latter. 

\begin{figure}
 \includegraphics[height=5.5cm]{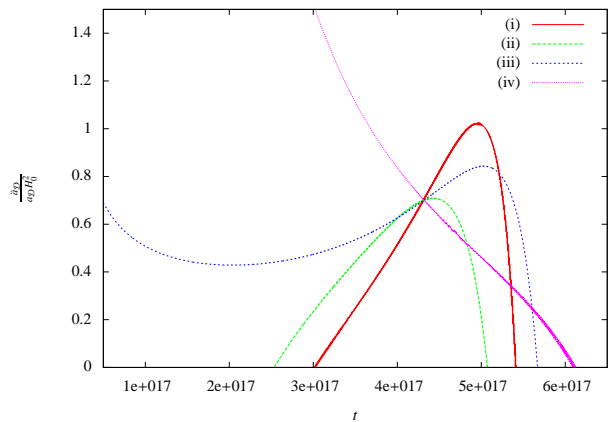} 
\caption{The dimensionless global acceleration parameter $\frac{\ddot{a}_{D}}{a_{D}H_{0}^{2}}$
is plotted versus time ($s$) as obtained through numerical integration, with the
`initial condition of $q_0= -0.7$. The values for the various 
parameters used are chosen to be the same as in the corresponding plots
of Fig.1.}
\label{fig2}
\end{figure}

\section{Conclusions}

To summarize, in this work we have explored the effect of backreaction
due to inhomogeneities on the evolution of the Universe undergoing present
acceleration. We have shown that the presence of the cosmic event horizon
causes the acceleration to slow down significantly with time. Our
results indicate the fascinating possibility of backreaction being
responsible not only for the present acceleration as shown in earlier
works \citep{buchert-3,rasanen-3}, but also leading
to a transition to another decelerated era in the future.
Another possibility following from our analysis is of the Universe
currently accelerating due to a different mechanism \citep{sahni,copeland},
but with backreaction \citep{buchert-1,buchert-2,buchert-3}  later causing acceleration
to slow down.  Our prediction of the future
slowing down of acceleration seems to
fit smoothly with the earlier era of structure formation and the
transition to acceleration in the standard $\Lambda$CDM model,
as shown here (transition red-shift $z_{T}\approx0.8$).

Before concluding, in context of the formalism used in the present work
it may be worthwhile to recapitulate some of the
present debate in the literature regarding averaging on a space like 
hypersurface \citep{buchert-1,buchert-2,buchert-3} as compared to taking
the average on the past light cone. The usefulness of the 
expansion rate averaged on any hypersurface is determined by relating it to 
observed quantities. It has been observed that the redshift and distance
can be expressed in terms of the average geometry alone, provided that the 
contribution of the null shear is negligible \citep{rasanen-25}. 
Observationally, the shear is known to be indeed small \citep{munshi}. 
Nonetheless, it has been claimed that neither averaging on a constant 
time hypersurface nor light cone averaging is easy to connect with the 
observations corresponding to parameters of the $\Lambda$CDM model \citep{lamb}.
The task of developing a procedure for light cone averaging is an ambitious
program and till date there is no standard formalism to do so. In a recent
paper three different types of light cone averaging have been proposed
\citep{gasp}, though much work remains to be done in
order to apply their technique to the problem of cosmic acceleration. 
On the other hand, our present work introduces an element of light cone 
physics from another perspective by considering an effect due 
to the event horizon.

Finally, it may be noted that though the event horizon is 
observer dependent, it follows from the symmetry of the equations
(\ref{eq:15}) and (\ref{voidfrac}) that our analysis leads to similar
conclusions for a ``void'' centric observer, as it does for a ``wall''
centric one. Note also that our analysis  is valid
while the event horizon exists. Hence, if the acceleration vanishes at some 
epoch in the future, one needs to consider backreaction without the 
event horizon beyond that epoch. The
scales which crossed ouside the horizon earlier, will begin re-entering with
backreaction from their associated inhomogeneities
impacting the evolution. Such a scenario
is somewhat reminiscent of the horizon crossing of modes during inflation in the
early universe, and their subsequent reentry with rich cosmological consequences. In the present context, the impact on the global evolution
of the reentering scales needs to be studied further. Moreover, it would
be worthwhile to investigate effects of the cosmic event horizon on
other models of backreaction in order to make more generic predictions
of observational interest.


\begin{thebibliography}{99}

\bibitem[\protect\citeauthoryear{Bolejko \& Andersson}{2008}]{bolejko} 
Bolejko K., and Andersson L., 2008, JCAP  10, 003 

\bibitem[\protect\citeauthoryear{Buchert}{2000}]{buchert-1} Buchert T., 2000,
Gen. Rel. Grav. 32, 105

\bibitem[\protect\citeauthoryear{Buchert \& Carfora }{2003}]{buchert-4}
Buchert T.,  and Carfora M., 2003, Phys. Rev. Lett. 90,
031101 

\bibitem[\protect\citeauthoryear{Buchert \& Carfora }{2008}]{buchert-2} Buchert
T.,  and Carfora M., 2008, Class. Quant. Grav. 25, 195001 

\bibitem[\protect\citeauthoryear{Copeland et al.}{2006}]{copeland} Copeland
E. J., Sami M., Tsujikawa S., 2006, Int. J. Mod. Phys. D 15, 1753

\bibitem[\protect\citeauthoryear{Gasperini et al.}{2009}]{gasperini} Gasperini
M., Marozzi G., and Veneziano G., 2009, JCAP 0903, 011

\bibitem[\protect\citeauthoryear{Gasperini et al.}{2011}]{gasp}
Gasperini M.,  Marozzi G., Nugier F., and Veneziano G., 2011, JCAP 07, 008

\bibitem[\protect\citeauthoryear{Hicken}{2009}]{hicken}
Hicken M., et al., 2009, APJ 700, 1097 

\bibitem[\protect\citeauthoryear{Ishibashi \& Wald}{2006}]{wald} Ishibashi A.,
and Wald R. M, 2006, Class. Quant. Grav. 23,
235 

\bibitem[\protect\citeauthoryear{Kolb et al.}{2005}]{kolb-1}
Kolb E. W., 
Matarrese S., Notari A., and Riotto A., 2005,
Phys. Rev. D 71, 023524 

\bibitem[\protect\citeauthoryear{Kolb et al.}{2008}]{kolb-2}
Kolb E. W., Marra S., Matarrese S., 2008, Phys. Rev. D
78, 103002 

\bibitem[\protect\citeauthoryear{Kolb \& Lamb}{2009}]{lamb}
Kolb E. W., and Lamb C. R., 2009, arXiv: 0911.3852

\bibitem[\protect\citeauthoryear{Kowalski}{2008}]{kowalski}
Kowalski M., et al., 2008, APJ 686, 749 

\bibitem[\protect\citeauthoryear{Mattson \& Mattson}{2010}]{mattson} Mattsson
M., and Mattsson T., 2010, JCAP 10, 021 

\bibitem[\protect\citeauthoryear{Melchiorri et al.}{2007}]{melchiorri} 
Melchiorri A., Pagano L., and Pandolfi S., 2007, Phys.
Rev. D 76, 041301(R) 

\bibitem[\protect\citeauthoryear{Munshi et al.}{2008}]{munshi} 
Munshi D., et al., 2008, Phys. Rep. 462, 67

\bibitem[\protect\citeauthoryear{Paranjape \& Singh}{2008}]{paranjape-1} 
Paranjape A., and Singh T. P., 2008,  Phys. Rev. D
78, 063522 

\bibitem[\protect\citeauthoryear{Perlmutter}{1998}]{perlmutter} 
Perlmutter S., et. al, 1998, Nature 391, 51 

\bibitem[\protect\citeauthoryear{Rasanen}{2004}]{rasanen-1} Rasanen S., 2004,
JCAP 0402, 003 

\bibitem[\protect\citeauthoryear{Rasanen}{2008}]{rasanen-2} Rasanen S., 2008,
JCAP 0804, 026 

\bibitem[\protect\citeauthoryear{Rasanen}{2009}]{rasanen-25} Rasanen S., 2009,
JCAP 02, 011 

\bibitem[\protect\citeauthoryear{Rasanen}{2010}]{rasanen-3} Rasanen S., 2010,
 Phys. Rev. D 81, 103512

\bibitem[\protect\citeauthoryear{Sahni}{2004}]{sahni} Sahni V., 2004, Lect. 
Notes Phys. 653, 141 

\bibitem[\protect\citeauthoryear{Seikel \& Schwarz}{2009}]{schwarz} 
Seikel M., and Schwarz D. J., 2009, JCAP 02, 024 

\bibitem[\protect\citeauthoryear{Singh}{2011}]{singh} Singh T. P., 2011,
preprint arXiv:1105.3450

\bibitem[\protect\citeauthoryear{Wiegand \& Buchert }{2010}]{buchert-3}
 Wiegand A.,  and Buchert T., 2010, Phys. Rev. D 82,
023523 

\bibitem[\protect\citeauthoryear{Weinberg}{1989}]{weinberg} Weinberg S., 1989,
Rev. Mod. Phys. 61, 1 

\bibitem[\protect\citeauthoryear{Wiltshire}{2007a}]{wiltshire-1} Wiltshire
D. L., 2007a, New J. Phys. 9, 377

\bibitem[\protect\citeauthoryear{Wiltshire}{2007b}]{wiltshire-2}
Wiltshire D. L., 2007b,  Phys. Rev. Lett. 99,
251101 

\bibitem[\protect\citeauthoryear{Zalaletdinov}{1992}]{zalaletdinov-1} 
 Zalaletdinov R., 1992, Gen. Rel. Grav. 24,
1015  


\end{thebibliography}
\end{document}